\documentclass[conference,a4,comsoc]{IEEEtran}
\usepackage{graphicx}
\usepackage{amssymb}
\usepackage{amsmath}
\usepackage{cite}
\usepackage{mathrsfs}
\usepackage[displaymath,mathlines]{lineno}
\usepackage{color}
\usepackage{tabulary}
\usepackage{multirow}
\usepackage{algpseudocode}
\usepackage{pbox}
\usepackage{multicol}
\usepackage{lipsum}
\usepackage{algorithm}
\usepackage{algpseudocode}
\usepackage{amsthm}
\usepackage{relsize}
\usepackage{lipsum}
\usepackage{epstopdf}
\usepackage{times}
\usepackage{enumitem}
\usepackage{float}
\usepackage{mathtools}
\usepackage{calc}
\usepackage{makecell}
\usepackage{caption}
\usepackage{subcaption}
\usepackage{soul}
\usepackage{algorithm}
\usepackage{fancyhdr}
\usepackage{verbatim}
\makeatletter
\newcommand{\doublewidetilde}[1]{{%
		\mathpalette\double@widetilde{#1}}}
\newcommand{\double@widetilde}[2]{%
		\sbox\z@{$\m@th#1\widetilde{#2}$}%
		\ht\z@=.5\ht\z@
		\widetilde{\box\z@}}
\makeatother

\allowdisplaybreaks

\makeatother
\makeatletter

\usepackage{titlesec}

\begin{document}
\addtolength{\topmargin}{0.03in}
\addtolength{\textheight}{-0.07in}

\title{\huge Energy-Efficient UAV Covert Communications via Metaheuristic Optimization \vspace{-0.25cm}
}
\author{Nguyen Minh Quan, Tran Thanh Thu, Nguyen The Ngoc Khue, Ninh Thi Thanh Tam, Trinh Van Chien\\
\fontsize{9}{9}{\textit{School of Information and Communications Technology, Hanoi University of Science and Technology, Vietnam}}\\
}
\vspace{-0.5cm}

\maketitle


\begin{abstract}

Wireless communication systems increasingly require not only confidentiality but also transmission undetectability, motivating covert communications. This paper studies an unmanned aerial vehicle (UAV) enabled covert communication system where a UAV serves a legitimate receiver in the presence of multiple active and colluding wardens with imperfect location information. We maximize covert energy efficiency (CEE) by jointly optimizing the UAV's 3D trajectory and transmit power while accounting for energy. To address the resulting non-convex problem with robust covertness constraints, we develop Genetic Algorithm (GA) and Particle Swarm Optimization (PSO)-based metaheuristic algorithms and compare them with a Block Coordinate Descent with Successive Convex Approximation (BCD-SCA) benchmark. Numerical results show that the proposed methods improve CEE while satisfying covertness constraints.

\end{abstract}


\section{Introduction}

Unmanned aerial vehicles have emerged as a promising platform for future wireless communication networks due to their high mobility, flexible deployment, and reliable line-of-sight (LoS) links with ground terminals~\cite{pu2026movable}. Compared with terrestrial infrastructure, UAV-assisted systems can be rapidly deployed in areas where fixed base stations are unavailable, damaged, or costly to install. Therefore, UAVs have been widely studied for emergency communications, coverage enhancement, data collection, relaying, and Internet-of-Things (IoT) connectivity~\cite{fan2026air}. Despite these advantages, UAV-enabled communication faces important security and privacy challenges. Due to the broadcast nature of wireless channels and the high altitude of UAV platforms, legitimate transmissions may be overheard or detected by unintended ground nodes. In critical scenarios such as military operations, disaster response, surveillance, or confidential sensing, protecting only the message content is insufficient~\cite{al2026uav}. It is also necessary to hide the existence of the transmission itself, which motivates covert communication, also known as low-probability-of-detection communication~\cite{wu2026uav}. A key challenge in UAV-enabled covert communication is the limited onboard energy. Unlike ground nodes, a UAV consumes energy for both wireless transmission and propulsion, including hovering and maneuvering. Since propulsion energy often dominates communication energy, trajectory design plays a crucial role in the overall system performance~\cite{santoso2026ai, thu2026energy}. The problem becomes even more challenging in adversarial environments with active and colluding wardens. Active wardens may transmit jamming signals to degrade the legitimate link, while colluding wardens can jointly process their observations to improve detection capability. Moreover, imperfect warden location information caused by sensing errors, outdated estimates, or adversarial mobility may lead trajectory and power control strategies to violate worst-case covertness constraints~\cite{ahmed2025toward}. Existing studies on UAV covert communication have mainly focused on covert throughput maximization through joint trajectory and transmit power optimization~\cite{wei2026uav}. Other works have considered energy-efficient UAV communication, but often without explicitly accounting for adversarial detection constraints or active colluding wardens~\cite{yang2024uav}. Consequently, the joint optimization of UAV trajectory and transmit power for covert energy efficiency under active colluding wardens and uncertain warden locations remains insufficiently explored. This motivates the need for a robust optimization framework that can simultaneously handle mobility constraints, propulsion energy consumption, transmit power allocation, and worst-case covertness requirements.

To address this gap, this paper investigates robust covert energy efficiency maximization for a UAV-enabled covert communication system. The main contributions include: (i) We formulate a robust covert energy efficiency maximization problem for UAV-enabled covert communication in the presence of active and colluding wardens with uncertain locations. (ii) We derive a deterministic reformulation of the worst-case covertness constraint based on the aggregate received power at the wardens, enabling tractable feasibility checking during optimization. (iii) We develop GA and PSO optimization algorithms for jointly optimizing the UAV trajectory and transmit power under mobility, energy, and covertness constraints. (iv) We compare the proposed metaheuristic approaches with a BCD-SCA benchmark and provide numerical results showing their effectiveness in improving covert energy efficiency while maintaining robust covertness.


\section{System Model}

\subsection{UAV Covert Model}
\begin{figure}[t]
    \centering
    \includegraphics[trim={0cm 1cm 0cm 1cm}, clip, width=\columnwidth]{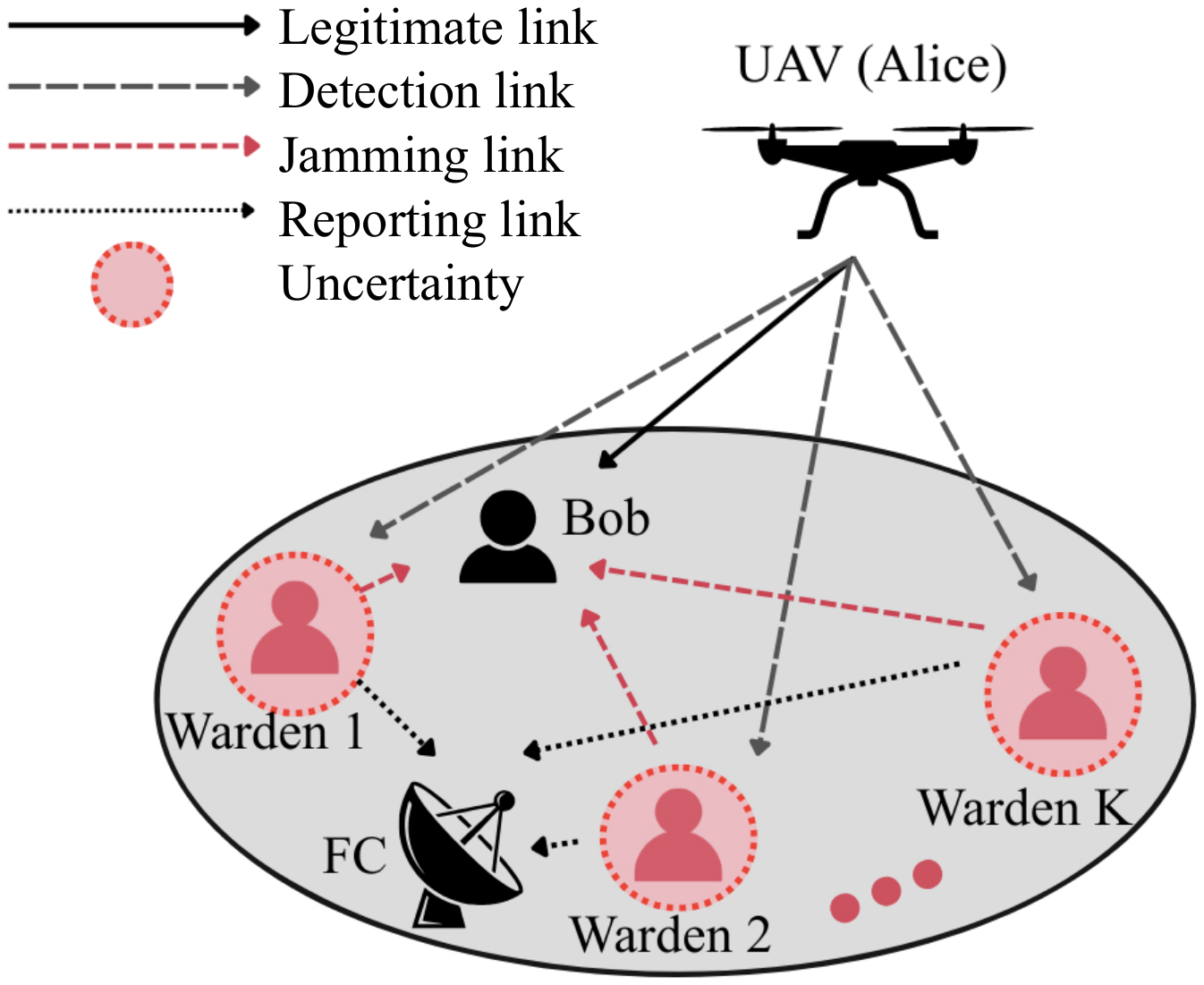}
    \caption{UAV covert communication with active wardens.}
    \label{fig:system_model}
\end{figure}

We consider the robust UAV-enabled covert communication system shown in Fig.~\ref{fig:system_model}, where a UAV (Alice) communicates with a ground receiver (Bob) in the presence of $K$ active wardens. Alice and Bob are each equipped with a single antenna, whereas every warden employs two antennas for simultaneous signal detection and jamming. In addition, the wardens report their local detection results to the Fusion Center (FC) through reporting links, where the received decisions are fused to determine whether the UAV is transmitting. A three-dimensional Cartesian coordinate system is adopted. Bob's horizontal location is denoted by 
$\mathbf{b}\triangleq[b_x,b_y]^{\mathsf T}\in\mathbb{R}^{2}$. The horizontal location of warden $k$ is \textcolor{black}{$\mathbf{w}_k\triangleq[w_{k,x},w_{k,y}]^{\mathsf T}$}, which is imperfectly known at the UAV.  Following the worst-case uncertainty model as 
\textcolor{black}{$\mathbf{w}_k \triangleq \hat{\mathbf{w}}_k+\Delta\mathbf{w}_k,\quad \|\Delta\mathbf{w}_k\|^2\le\epsilon_w^2,\;\forall k,$} 
with \textcolor{black}{$\hat{\mathbf{w}}_k$} is the estimated location, \textcolor{black}{$\Delta\mathbf{w}_k$} is the estimation error, and $\epsilon_w$ denotes the uncertainty radius.

The mission duration $T$ is equally divided into $N$ time slots with $\Delta t=T/N$. The UAV position in slot $n$ is represented by the horizontal coordinate $\mathbf{q}[n]$ and altitude $H[n]$. The UAV mobility constraints are:
\begin{align}
        \|\mathbf{q}[n\!+\!1]-\mathbf{q}[n]\| &\le V_h\,\Delta t,\quad \forall n, \label{eq2} \\ 
        |H[n\!+\!1]-H[n]| &\le V_v\,\Delta t,\quad \forall n, \label{eq3} \\
        H_{\min}\le H[n] &\le H_{\max},\quad \forall n,  \label{eq4} \\
        \mathbf{q}[0]=\mathbf{q}_{\mathrm{I}},\ \mathbf{q}[N]=\mathbf{q}_{\mathrm{F}},\quad
        &H[0]=H_{\mathrm{I}},\ H[N]=H_{\mathrm{F}}.
        \label{eq5} 
    \end{align}
The UAV travel distance during slot $n$ is:
\begin{equation}
\mathcal{D}_n=
\sqrt{\|\mathbf{q}[n+1]-\mathbf{q}[n]\|^2+\left(H[n+1]-H[n]\right)^2}.
\label{eq:3}
\end{equation}

\subsection{Channel Model}
The large-scale channel power gains from the UAV to Bob and to the sensing antenna of warden $k$ in time slot $n$ are modeled as
    \begin{align}
    g_{ab}[n] &= \frac{\kappa_0}{\|\mathbf{q}[n]-\mathbf{b}\|^2 + H^2[n]}, \quad \forall n, \label{eq:LoS_uav_bob}\\
    g_{ak}[n] &= \frac{\kappa_0}{\|\mathbf{q}[n]-\mathbf{w}_k\|^2 + H^2[n]}, \quad \forall n,\forall k, \label{eq:LoS_uav_warden}
    \end{align}
where $\kappa_0$ denotes the channel power gain at a reference distance of $1$m. The corresponding complex channel coefficients can be written as
    \begin{equation}
    \begin{aligned}
    h_{ab}[n] &= \sqrt{g_{ab}[n]}\,e^{j\phi_{ab}[n]}, h_{ak}[n] = \sqrt{g_{ak}[n]}\,e^{j\phi_{ak}[n]},
    \end{aligned}
    \end{equation}
where $\phi_{ab}[n]$ and $\phi_{ak}[n]$ denote the LoS phases, which do not affect the rate expression in our Single-Input Single-Output (SISO) setting and are thus omitted in the sequel. The complex channel coefficient in slot $n$ is modeled as
    \begin{equation}
    \begin{aligned}
    h_{kb}[n] = \sqrt{\kappa_0\, d_{kb}^{-\alpha}}\; g_{kb}[n], \quad \forall n,k,
    \end{aligned}
    \label{eq:chan_jam_bob}
    \end{equation}
where $\alpha$ is the path-loss exponent, $d_{kb}\triangleq \|\mathbf{b}-\mathbf{w}_k\|$ is the horizontal distance between Bob and warden $k$, and $g_{kb}[n]\sim \mathcal{CN}(0,1)$ denotes small-scale fading. Since the exact warden location $\mathbf{w}_k$ is imperfectly known, we use a conservative lower bound on $d_{kb}$ based on the triangle inequality. 
\textcolor{black}{Specially for $\mathbf{w}_k=\hat{\mathbf{w}}_k+\Delta\mathbf{w}_k$}
with
\textcolor{black}{$\|\Delta\mathbf{w}_k\|\le\epsilon_w$},
the distance $d_{kb}$ is lower-bounded as
{\color{black}
\begin{equation}
\begin{aligned}
d_{kb}
&=\|\mathbf{b}-\mathbf{w}_k\| \\
&\geq \|\mathbf{b}-\hat{\mathbf{w}}_k\|
      -\|\Delta\mathbf{w}_k\| \\
&\geq \|\mathbf{b}-\hat{\mathbf{w}}_k\|-\epsilon_w
\triangleq d_{kb}^{\mathrm{lb}}, \\
\underline{d}_{kb}
&\triangleq \max\left\{d_{kb}^{\mathrm{lb}},\,d_{\min}\right\}.
\end{aligned}
\label{eq:d_kb_lower}
\end{equation}}
where $d_{\min}>0$ is introduced to avoid distance singularity.
This conservative distance bound follows the worst-case robust optimization methodology and will be used to obtain a tractable lower bound on Bob's achievable rate in the presence of active jamming.

\subsection{Transmit Power Constraint}

Let $P[n]$ denote the UAV transmit power in slot $n$. It satisfies the peak and average power limitations dictated by the hardware constraints:
    \begin{equation}
        0\le P[n]\le P_{\max},\quad \frac{1}{N}\sum_{n=1}^{N} P[n]\le \bar{p},
        \label{eq:power_constraint}
    \end{equation}
where $p_{\max}$ and $\bar{p}$ represent the maximum peak power and average power budget, respectively. Each warden transmits a malicious jamming signal with a maximum power $p_J$ to actively disrupt the legitimate transmission. The corresponding jamming symbol emitted from warden $k$ in slot $n$ is denoted by $s_k[n]$ with $\mathbb{E}[|s_k[n]|^2]=1$, with $\mathbb{E}[\cdot]$ being the expectation operator. Let $x[n]$ be the UAV information-bearing symbol directed to Bob, satisfying $\mathbb{E}[|x[n]|^2]=1$. By considering the simultaneous transmissions from the aerial node and the colluding ground network, the aggregate received signal at Bob in slot $n$ is expressed as:
    \begin{equation}
        y_b[n] = \sqrt{P[n]}\, h_{ab}[n]\, x[n] + \sum_{k=1}^{K}\sqrt{p_J}\, h_{kb}[n]\, s_k[n] + v_b[n],
        \label{eq:rx_bob}
    \end{equation}
where $v_b[n]\sim \mathcal{CN}(0,\sigma_b^2)$ denotes the additive white Gaussian noise (AWGN) at Bob. The instantaneous achievable rate (in bps/Hz) at Bob in slot $n$ is given by:
    \begin{equation}
        R_b[n]=\log_2\!\left(1+\frac{P[n]\, |h_{ab}[n]|^2}{p_J\sum_{k=1}^{K} |h_{kb}[n]|^2+\sigma_b^2}\right),\quad \forall n.
        \label{eq:rate_inst}
    \end{equation}

\subsection{Lower-Bound Rate via Jensen's Inequality}
The interference caused by the active wardens is random due to the Rayleigh fading of the warden-Bob links. By substituting \eqref{eq:chan_jam_bob} into \eqref{eq:rate_inst}, the instantaneous achievable rate at Bob in time slot $n$ as
\begin{equation}
R_b[n]
=
\log_2\!\left(
1+
\frac{P[n]g_{ab}[n]}
{p_J\kappa_0\sum_{k=1}^{K} d_{kb}^{-\alpha}|g_{kb}[n]|^2+\sigma_b^2}
\right).
\label{eq:rate_inst_simplified}
\end{equation}
Define the aggregate jamming interference at Bob as $\mathcal I[n]
\triangleq
p_J\kappa_0
\sum_{k=1}^{K} d_{kb}^{-\alpha}|g_{kb}[n]|^2.$
For fixed $P[n]$ and $g_{ab}[n]$, let
$f(x)=\log_2\left(1+\frac{P[n]g_{ab}[n]}{x+\sigma_b^2}\right), x\ge 0.$
It can be verified that $f''(x)>0$, and hence $f(x)$ is convex in $x$. Therefore, by Jensen's inequality,
\begin{equation}
\begin{aligned}
\mathbb E[R_b[n]]
&=
\mathbb E\!\left[
\log_2\left(
1+\frac{P[n]g_{ab}[n]}{\mathcal I[n]+\sigma_b^2}
\right)
\right]  \\
&\ge
\log_2\left(
1+\frac{P[n]g_{ab}[n]}
{\mathbb E[\mathcal I[n]]+\sigma_b^2}
\right).
\end{aligned}
\label{eq:jensen_step_simplified}
\end{equation}
Since $\mathbb E[|g_{kb}[n]|^2]=1$, the mean aggregate jamming interference is
$\mathbb E[\mathcal I[n]]
=
p_J\kappa_0
\sum_{k=1}^{K} d_{kb}^{-\alpha}.$ \textcolor{black}{By applying the conservative distance lower bound in ${d}_{kb}$,
the mean aggregate jamming interference is upper-bounded as
$\mathbb{E}[\mathcal{I}[n]]
\leq
p_J\kappa_0
\sum_{k=1}^{K}
\underline{d}_{kb}^{-\alpha}.$
}
Because $f(x)$ is monotonically decreasing in $x$, replacing $\mathbb E[\mathcal I[n]]$ by the upper bound yields the following conservative deterministic lower bound:
\begin{equation}
\mathbb E[R_b[n]]
\ge
\log_2\!\left(
1+
\frac{P[n]g_{ab}[n]}{\Gamma}
\right)
\triangleq
R_b^{\mathrm{LB}}[n],
\label{eq:rate_lb}
\end{equation}
where $\Gamma
\triangleq
p_J\kappa_0
\sum_{k=1}^{K}
\underline d_{kb}^{-\alpha}
+
\sigma_b^2$ is the conservative effective interference-plus-noise term. The lower-bound rate $R_b^{\mathrm{LB}}[n]$ is then used as the communication utility in the subsequent energy-efficiency maximization.




\subsection{Covert Communication Constraint}

The likelihood ratio test can be expressed as:

\begin{equation}
\label{eq:likelihood_ratio}
    \frac{\prod_{l=1}^{L} f\!\left(Y_{FC}^{(l)}[n] \mid \mathcal{H}_1\right)} {\prod_{l=1}^{L} f\!\left(Y_{FC}^{(l)}[n] \mid \mathcal{H}_0\right)} \underset{\mathcal{D}_0}{\overset{\mathcal{D}_1}{\gtrless}} 1,
\end{equation}
where $f(Y_{\mathrm{FC}}^{(l)}[n] \mid \mathcal{H}_0) = \mathcal{CN}(0, \mu_1[n])$ and 
$f(Y_{\mathrm{FC}}^{(l)}[n] \mid \mathcal{H}_1) = \mathcal{CN}(0, \mu_1[n] + \mu_2[n])$ \textcolor{black}{with $\mu_1[n] = \sum_{k=1}^{K}\sigma_k^2[n]$, and  $\mu_2[n] = P[n]\sum_{k=1}^{K}g_{ak}[n]$.} 
By applying the probability density function of a circularly symmetric complex normal distribution, the products 
$\prod_{l=1}^{L} f(Y_{\mathrm{FC}}^{(l)}[n] \mid \mathcal{H}_0)$ and 
$\prod_{l=1}^{L} f(Y_{\mathrm{FC}}^{(l)}[n] \mid \mathcal{H}_1)$ 
can be rewritten as follows \cite{yang2024uav}:

\begin{equation}
\begin{aligned}
&\prod_{l=1}^{L} f(Y_{\mathrm{FC}}^{(l)}[n] \mid \mathcal{H}_0) = \frac{1}{(\pi \mu_1[n])^L} e^{\left( -\frac{\sum_{l=1}^{L} |Y_{\mathrm{FC}}^{(l)}[n]|^2}{\mu_1[n]} \right)},
\end{aligned}
\end{equation}

\begin{equation}
\begin{aligned}
&\prod_{l=1}^{L} f(Y_{\mathrm{FC}}^{(l)}[n] \mid \mathcal{H}_1) = \frac{1}{(\pi (\mu_1[n] + \mu_2[n]))^L} e^{\left( -\frac{\sum_{l=1}^{L} |Y_{\mathrm{FC}}^{(l)}[n]|^2}{\mu_1[n] + \mu_2[n]} \right)}.
\end{aligned}
\end{equation}

Then, (\ref{eq:likelihood_ratio}) can be further rewritten as:

\begin{equation}
\label{eq:lh_rate}
     \frac{(\pi(\mu_1[n] + \mu_2[n]))^{-L} \exp\!\left(-\frac{\sum_{l=1}^{L} \left|Y_{FC}^{(l)}[n]\right|^2}{\mu_1[n] + \mu_2[n]}\right)} {(\pi \mu_1[n])^{-L} \exp\!\left(-\frac{\sum_{l=1}^{L} \left|Y_{FC}^{(l)}[n]\right|^2}{\mu_1[n]}\right)} \underset{\mathcal{D}_0}{\overset{\mathcal{D}_1}{\gtrless}} 1.
\end{equation}

Applying the natural logarithm, (\ref{eq:lh_rate}) can be expressed as:
\begin{equation}
\begin{aligned}
&L \log\left( \frac{\mu_1[n]}{\mu_1[n] + \mu_2[n]} \right) \\
&+ \sum_{l=1}^{L} |Y_{\mathrm{FC}}^{(l)}[n]|^2 \left( \frac{1}{\mu_1[n]} - \frac{1}{\mu_1[n] + \mu_2[n]} \right) \underset{\mathcal{D}_0}{\overset{\mathcal{D}_1}{\gtrless}} 0,
\end{aligned}
\label{eq:likelihood_expanded}
\end{equation}
which can be simplified as $\sum_{l=1}^{L} \left|Y_{FC}^{(l)}[n]\right|^2 \underset{\mathcal{D}_0}{\overset{\mathcal{D}_1}{\gtrless}} L\Delta t^*[n]$, with $\Delta t^*[n] = \frac{\mu_1[n]\bigl(\mu_1[n] + \mu_2[n]\bigr)}{\mu_2[n]} \log\!\left(\frac{\mu_1[n] + \mu_2[n]}{\mu_1[n]}\right)$. Since $Y_{FC}^{(l)}[n]$ follows a circularly symmetric complex normal distribution, the sum $\sum_{l=1}^{L} \left| Y_{FC}^{(l)}[n] \right|^2$ follows Gamma distribution, where: $Y_{FC}^{(l)}[n] \mid \mathcal{H}_0 \sim \Gamma(L, \mu_1[n])$ and $Y_{FC}^{(l)}[n] \mid \mathcal{H}_1 \sim \Gamma(L, \mu_1[n] + \mu_2[n])$. Therefore, the false alarm probability and miss detection can be expressed as follow:
\begin{align}
     &P_{FA}[n] = \Pr(\mathcal{D}_1 \mid \mathcal{H}_0) = \Pr\left(\sum_{l=1}^{L} \left|Y_{FC}^{(l)}[n]\right|^2 > L\Delta t^*[n] \mid \mathcal{H}_0\right) \nonumber \\
     &=1 - \Pr\left(\sum_{l=1}^{L} \left|Y_{FC}^{(l)}[n]\right|^2 \leq L\Delta t^*[n] \mid \mathcal{H}_0\right) \nonumber \\
     &=1 - \int_0^{L\Delta t^*[n]}{\frac{x^{L - 1}e^{-\frac{x}{\mu_1[n]}}}{\Gamma(L)\mu_1[n]^L}}dx
     = 1 - \frac{\gamma(L, \frac{L\Delta t^*[n]}{\mu_1[n]})}{\Gamma(L)}. 
\end{align}
{\color{black}
\begin{align}
    P_{MD}[n] &= \Pr(\mathcal{D}_0 \mid \mathcal{H}_1) = \Pr\left(\sum_{l=1}^{L} \left|Y_{FC}^{(l)}[n]\right|^2 < L\Delta t^*[n] \mid \mathcal{H}_1\right) \nonumber \\
    &= \frac{\gamma(L, \frac{L\Delta t^*[n]}{\mu_1[n] + \mu_2[n]})}{\Gamma(L)}. 
\end{align}
}
Based on these, the optimal detection error probability can be expressed as ~\cite{yang2024uav}: 

\begin{equation}
    \xi^*[n] = 1 - \frac{\gamma\!\left(L, \tfrac{L\Delta t^*[n]}{\mu_1[n]}\right)}{(L-1)!} + \frac{\gamma\!\left(L, \tfrac{L\Delta t^*[n]}{\mu_1[n] + \mu_2[n]}\right)}{(L-1)!}.
\end{equation}
To ensure covert communication, the following constraint must be satisfied $\xi^*[n] \geq 1 - \varepsilon, \forall n,$ where $\varepsilon \in [0, 1]$ describes the level of covertness, which is sufficiently small.



\section{Covert Energy Efficiency Maximization}
\label{sec:cee}



\subsection{Problem Formulation}
\label{subsec:formulation}

For ease of presentation, let
\textcolor{black}{$\mathbf{P}   = [P[0],\ldots,P[N]]^{\mathsf{T}}$,
$\mathbf{Q}   = [\mathbf{q}[0],\ldots,\mathbf{q}[N]]^{\mathsf{T}}$, and
$\mathbf{H}   = [H[0],\ldots,H[N]]^{\mathsf{T}}$}
denote the transmit power, the horizontal trajectory, and
the altitude profile of the UAV, respectively. The total energy consumed during the mission consists of two parts:
the propulsion energy required by the UAV and the
transmission energy consumed for information delivery.
Specifically, the propulsion power at time slot $n$ is modelled
as $P_{\mathrm{fly}}\bigl(v[n]\bigr)$ in \cite{van2025metaheuristic}.

The total energy consumed over the mission horizon~$T$ is
\begin{align}
  E_{\mathrm{total}}(\mathbf{P},\mathbf{Q},\mathbf{H})
  = \Delta t\sum_{n=0}^{N}
    \Bigl[P_{\mathrm{fly}}\bigl(v[n]\bigr) + P[n]\Bigr].
  \label{eq:total_energy}
\end{align}
The covert energy efficiency is defined as:
\begin{align}
  \Psi_{\mathrm{CEE}}(\mathbf{P},\mathbf{Q},\mathbf{H})
  = \frac{\Delta t\displaystyle\sum_{n=0}^{N} R_{b}^{\mathrm{LB}}[n]}
         {E_{\mathrm{total}}(\mathbf{P},\mathbf{Q},\mathbf{H})}
  \quad [\text{bits/Joule/Hz}],
  \label{eq:CEE}
\end{align}
The CEE maximization problem is stated as
\begin{subequations}\label{P1}
\begin{align}
  \mathbf{P1}:\quad
  &\max_{\mathbf{P},\mathbf{Q},\mathbf{H}}\;
    \Psi_{\mathrm{CEE}}(\mathbf{P},\mathbf{Q},\mathbf{H})
  \label{P1:obj} \\[3pt]
  \mathrm{s.t.}\quad 
  & \eqref{eq2}, \eqref{eq3}, \eqref{eq4}, \eqref{eq5}, \eqref{eq:power_constraint} \\
  &P[n]\sum_{k=1}^{K}g_{ak}[n] \leq \mu_{2}^{\max}[n],
    \quad \forall n,
  \label{P1:covert}
\end{align}
\end{subequations}
where ~\eqref{P1:covert} \textcolor{black}{includes
$\mu_{2}^{\max}[n]$ is the maximum allowable aggregate received power at the wardens, obtained via bisection
such that $\xi^{*}[n]\geq 1-\varepsilon$.}

\section{Proposed Methods}
To solve the non-convex CEE maximization problem, we investigate two population-based metaheuristic algorithms, namely GA and PSO \cite{van2026single}. For a fair comparison, both methods use the same computational budget, i.e., $N_{\mathrm{pop}}\times N_{\mathrm{gen}}$, and are compared with the BCD-SCA benchmark. In both GA and PSO, each candidate solution, denoted by $\mathbf{\Xi}$, is represented as a real-valued vector encoding the UAV trajectory, altitude, and transmit power, with dimension $4N-6$.

\textit{Genetic Algorithm:} The GA evolves a population of $N_{\mathrm{pop}}$ chromosomes over $N_{\mathrm{gen}}$ generations to maximize the proposed CEE objective. The main genetic operations are summarized as follows.

\begin{itemize}
    \item \textbf{Selection}: Tournament selection with a tournament size of four is employed to choose parent chromosomes.
    
    \item \textbf{Crossover}: Single-point crossover is performed with probability $p_c$, where two parents exchange genetic segments after a randomly selected crossover point.
    
    \item \textbf{Mutation}: Each gene is independently mutated with probability $p_m$ to maintain population diversity. The waypoint coordinates are perturbed by $\mathcal{U}(-\delta_{xy},\delta_{xy})$, whereas the transmit power genes are perturbed by $\mathcal{U}(-\delta_P,\delta_P)$.
    
    \item \textbf{Elitism}: The best $N_{\mathrm{pop}}$ individuals from the combined parent-offspring population are retained for the next generation, ensuring that the current best solution is preserved.
\end{itemize}

\textit{Particle Swarm Optimization:} The PSO algorithm maintains a swarm of $N_{\mathrm{pop}}$ particles, where each particle represents a candidate solution. During iteration $g$, particle $i$ updates its velocity and position according to

\begin{equation}
\begin{split}
\mathbf{v}_i^{(g+1)}
=&\,w\mathbf{v}_i^{(g)}
+c_1r_1(\mathbf{pbest}_i-\mathbf{x}_i^{(g)})+c_2r_2(\mathbf{gbest}-\mathbf{x}_i^{(g)}),
\end{split}
\label{eq:updateparticle}
\end{equation}
where $w$ denotes the inertia weight, $c_1$ and $c_2$ are the cognitive and social acceleration coefficients, respectively, and $r_1,r_2\sim\mathcal{U}(0,1)$. The particle position is then updated as
\begin{equation}
\mathbf{x}_i^{(g+1)}
=
\mathbf{x}_i^{(g)}
+
\mathbf{v}_i^{(g+1)}.
\label{eq:updateparpos}
\end{equation}

\textbf{Fitness Evaluation:} Since the proposed repair mechanism guarantees that every decoded solution satisfies the UAV mobility, transmit-power, and robust covertness constraints, the fitness function is directly defined as the covert energy efficiency (CEE), i.e.,
\begin{align}
    f(\mathbf{\Xi})
    =
    \Psi_{\mathrm{CEE}}
    \bigl(
    \mathbf{P}(\mathbf{\Xi}),
    \mathbf{Q}(\mathbf{\Xi}),
    \mathbf{H}(\mathbf{\Xi})
    \bigr).
    \label{eq:fitness}
\end{align}
For both GA and PSO, the fitness value is evaluated for every candidate solution in each generation (or iteration). After completing $N_{\mathrm{gen}}$ iterations, the individual with the highest fitness is selected as the final solution of the corresponding optimization algorithm.

\textit{Block Coordinate Descent with Successive Convex Approximation:} As a benchmark, the conventional BCD-SCA algorithm is adopted to solve the proposed CEE maximization problem. Unlike metaheuristic approaches, BCD-SCA iteratively updates the transmit power and UAV trajectory by alternately solving two optimization subproblems until the increase in CEE falls below a prescribed threshold $\epsilon_0$.

\textbf{Power Optimization Subproblem:} With the UAV trajectory $(\mathbf Q,\mathbf H)$ fixed, the propulsion power $P_{\mathrm{fly}}[n]$ is determined by the trajectory and hence
remains constant. From (26c), the covertness constraint yields the following covertness power limit:
\begin{align}
P_{\mathrm{cov}}[n]
=
\frac{\mu_2^{\max}[n]}
{\sum_{k=1}^{K}g_{ak}[n]},
\quad\forall n,
\label{eq:p_cov}
\end{align}
which leads to the effective power bound
$\tilde{P}_{\max}[n]
=
\min
\left(
P_{\max},
P_{\mathrm{cov}}[n]
\right).$ Accordingly, the transmit-power optimization is formulated as the
following fractional programming problem:
\begin{equation}
\max_{\mathbf P}
\frac{
\sum_{n=0}^{N}R_b^{\mathrm{LB}}[n]
}{
\sum_{n=0}^{N}
\left(P_{\mathrm{fly}}[n]+P[n]\right)
},
\end{equation}
subject to the average-power constraint and
$0\leq P[n]\leq\tilde P_{\max}[n]$. For a given parameter $\lambda$, the resulting
subproblem is
\begin{equation}
\max_{\mathbf P}
\left\{
\sum_{n=0}^{N}R_b^{\mathrm{LB}}[n]
-
\lambda
\sum_{n=0}^{N}
\left(P_{\mathrm{fly}}[n]+P[n]\right)
\right\}.
\end{equation}
The parameter $\lambda$ is iteratively updated as
\begin{equation}
\lambda
=
\frac{
\sum_{n=0}^{N}R_b^{\mathrm{LB}}[n]
}{
\sum_{n=0}^{N}
\left(P_{\mathrm{fly}}[n]+P[n]\right)
},
\end{equation}
until convergence.

\begin{figure*}[t]
    \centering
    \includegraphics[trim= {0cm 0cm 0cm 1cm}, clip, width=1\textwidth]{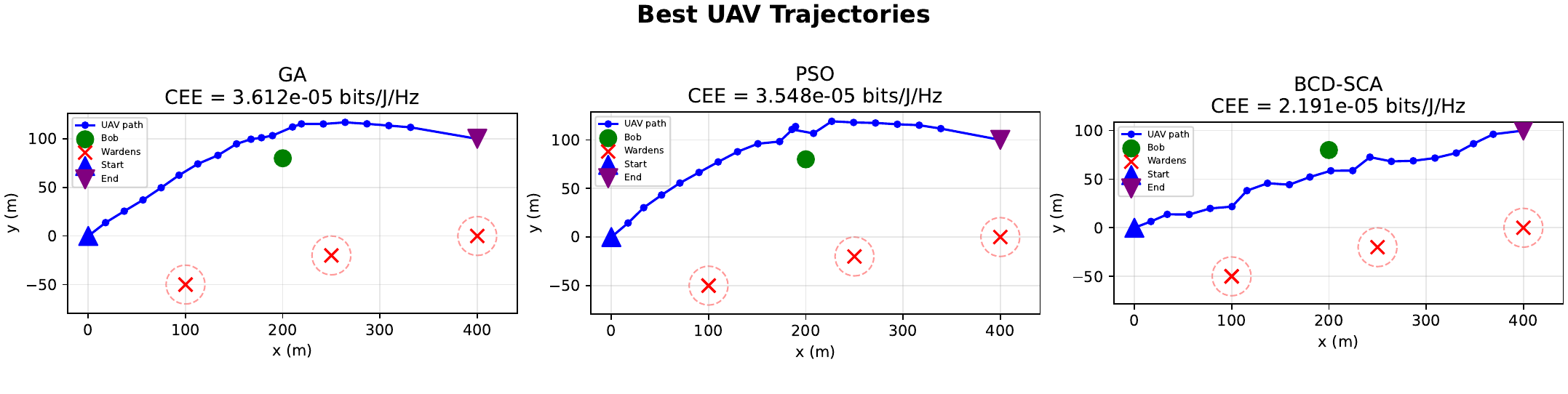}
    \caption{Optimized 2D horizontal trajectories of the UAV under different optimization schemes.}
    \label{fig:trajectory}
\end{figure*}

\textbf{Trajectory Optimization:} With the transmit power fixed, the UAV trajectory is optimized using the SCA technique to handle the remaining non-convex constraints. The CEE objective therefore becomes a function of $(\mathbf{Q},\mathbf{H})$ only.
We introduce slack variables
$\boldsymbol{\vartheta}=\{\vartheta[n]\}$ and
$\boldsymbol{\varsigma}_{k}=\{\varsigma_k[n]\}$, $\forall n,k$,
to decouple the trajectory from the channel gains, and recast the
trajectory subproblem as
\begin{subequations}\label{P1.2}
\begin{align}
  \mathbf{P1.2}:\quad
  &\max_{\mathbf{Q},\mathbf{H},
         \boldsymbol{\vartheta},
         \{\boldsymbol{\varsigma}_k\}}\;
    \frac{\Delta t\displaystyle\sum_{n=0}^{N}
          \log_2\!\left(1+\dfrac{P[n]\kappa_0}{\vartheta[n]}\right)}
         {E_{\mathrm{total}}}
  \label{P1.2:obj} \\[3pt]
  \mathrm{s.t.}\quad
  &\vartheta[n] \geq \|\mathbf{q}[n]-\mathbf{b}\|^{2}+H[n]^{2},
    \quad \forall n,
  \label{P1.2:theta} \\
  &\varsigma_k[n] \leq
    \|\mathbf{q}[n]-\hat{\mathbf{w}}_k\|^{2}+H[n]^{2},
    \quad \forall n,k,
  \label{P1.2:sigma} \\
  &P[n]\sum_{k=1}^{K}
    \frac{\kappa_0}{\varsigma_k[n]}
    \leq \mu_{2}^{\max}[n],
    \quad \forall n,
  \label{P1.2:covert_slack} \\
  &\varsigma_k[n] \geq H_{\min}^{2},
    \quad \forall n,k,
  \label{P1.2:sigma_lb} \\
  &\eqref{eq2}, \eqref{eq3}, \eqref{eq4}, \eqref{eq5}, \eqref{eq:power_constraint}
  \label{P1.2:mob}
\end{align}
\end{subequations}
Problem~\textbf{P1.2} remains non-convex due
to~\eqref{P1.2:theta} and~\eqref{P1.2:sigma}.
We apply the first-order Taylor
expansion around the $m$-th iterate
$(\mathbf{Q}^{(m)},\mathbf{H}^{(m)})$ to obtain the convex
surrogates
\begin{align}
\begin{aligned}
\|\mathbf{q}[n]-\mathbf{b}\|^{2}
&\ge \|\mathbf{q}^{(m)}[n]-\mathbf{b}\|^{2} + 2\bigl(\mathbf{q}^{(m)}[n]-\mathbf{b}\bigr)^{\dagger} \\
&\qquad \cdot \bigl(\mathbf{q}[n]-\mathbf{q}^{(m)}[n]\bigr),
\end{aligned}
\label{eq:taylor_q} \\
\begin{aligned}
H[n]^{2}
&\ge \bigl(H^{(m)}[n]\bigr)^{2}+ 2H^{(m)}[n]\bigl(H[n]-H^{(m)}[n]\bigr),
\end{aligned}
\label{eq:taylor_h}
\end{align}
which are substituted into~\eqref{P1.2:theta}.
Similarly, the S-procedure is applied to handle the robust warden-side constraint~\eqref{P1.2:sigma}, yielding the linear matrix inequality
\begin{align}
  \begin{pmatrix}
    \bigl(\phi[n]+1\bigr)\mathbf{I} &
    -\bigl(\mathbf{q}[n]-\hat{\mathbf{w}}_k\bigr) \\
    -\bigl(\mathbf{q}[n]-\hat{\mathbf{w}}_k\bigr)^{\dagger} &
    \tilde{\theta}_k[n] - \epsilon_w^2\phi[n]
  \end{pmatrix}
  \succeq \mathbf{0},
  \quad\forall n,k,
  \label{eq:lmi}
\end{align}
where
$\tilde{\theta}_k[n]
 = \|\mathbf{q}^{(m)}[n]-\hat{\mathbf{w}}_k\|^{2}
   +2(\mathbf{q}^{(m)}[n]-\hat{\mathbf{w}}_k)^{\dagger}
    (\mathbf{q}[n]-\mathbf{q}^{(m)}[n])
   +2H^{(m)}[n]H[n]
   -(H^{(m)}[n])^{2}
   -\varsigma_k[n]$
and $\phi[n]\geq 0$ is an auxiliary scalar.
The denominator $E_{\mathrm{total}}$ in~\eqref{P1.2:obj}
contains the non-convex propulsion term $P_{\mathrm{fly}}(v[n])$.
We approximate it by its first-order Taylor expansion around
$v^{(m)}[n]$:
\begin{align}
P_{\mathrm{fly}}\bigl(v[n]\bigr)
&\approx P_{\mathrm{fly}}\bigl(v^{(m)}[n]\bigr) + \dot{P}_{\mathrm{fly}}\bigl(v^{(m)}[n]\bigr)
\bigl(v[n]-v^{(m)}[n]\bigr) \nonumber \\
&\triangleq \tilde{P}_{\mathrm{fly}}^{(m)}\bigl(v[n]\bigr),
\label{eq:prop_sca}
\end{align}
Substituting \eqref{eq:taylor_q}--\eqref{eq:prop_sca} into Problem~\textbf{P1.2} yields a convex approximation, denoted as Problem~\textbf{P1.2c}. The resulting problem can be efficiently solved using standard convex optimization solvers such as CVX with MOSEK.

\section{Numerical Results}
This section compares the proposed GA and PSO algorithms with the BCD-SCA benchmark. The UAV flies from $\mathbf q_I=(0,0)$~m to $\mathbf q_F=(400,100)$~m, with the initial and final altitudes fixed at $100$~m. The altitude range is $[80,120]$~m, and the maximum horizontal and vertical speeds are both $10$~m/s. Bob is located at $\mathbf b=(200,80)$~m, while three active wardens are located at $\mathbf w_1=(100,-50)$~m, $\mathbf w_2=(250,-20)$~m, and $\mathbf w_3=(400,0)$~m. The main parameters are $\alpha=2.6$, $\kappa_0=-60$~dB, $\sigma_b^2=\sigma_k^2=-110$~dBm, $P_{\max}=20$~dBm, $\bar P=P_{\max}/4$, $T=40$~s, $N=20$, $L=10$, $\epsilon_w = 20$~m, and $\varepsilon=0.3$. For GA and PSO, the population/swarm size is $N_{\mathrm{pop}}=50$, and the maximum number of generations/iterations is $N_{\mathrm{gen}}=100$. The GA uses $p_c=0.8$ and $p_m=0.1$, while the PSO uses $w=0.7$ and $c_1=c_2=1.5$. The algorithms are evaluated in terms of covert energy efficiency, trajectory design, power allocation, convergence behavior, and robustness to key system parameters.

\subsection{Trajectory and Altitude Optimization}

\begin{figure}[t]
    \centering
    \includegraphics[width=0.95\linewidth]{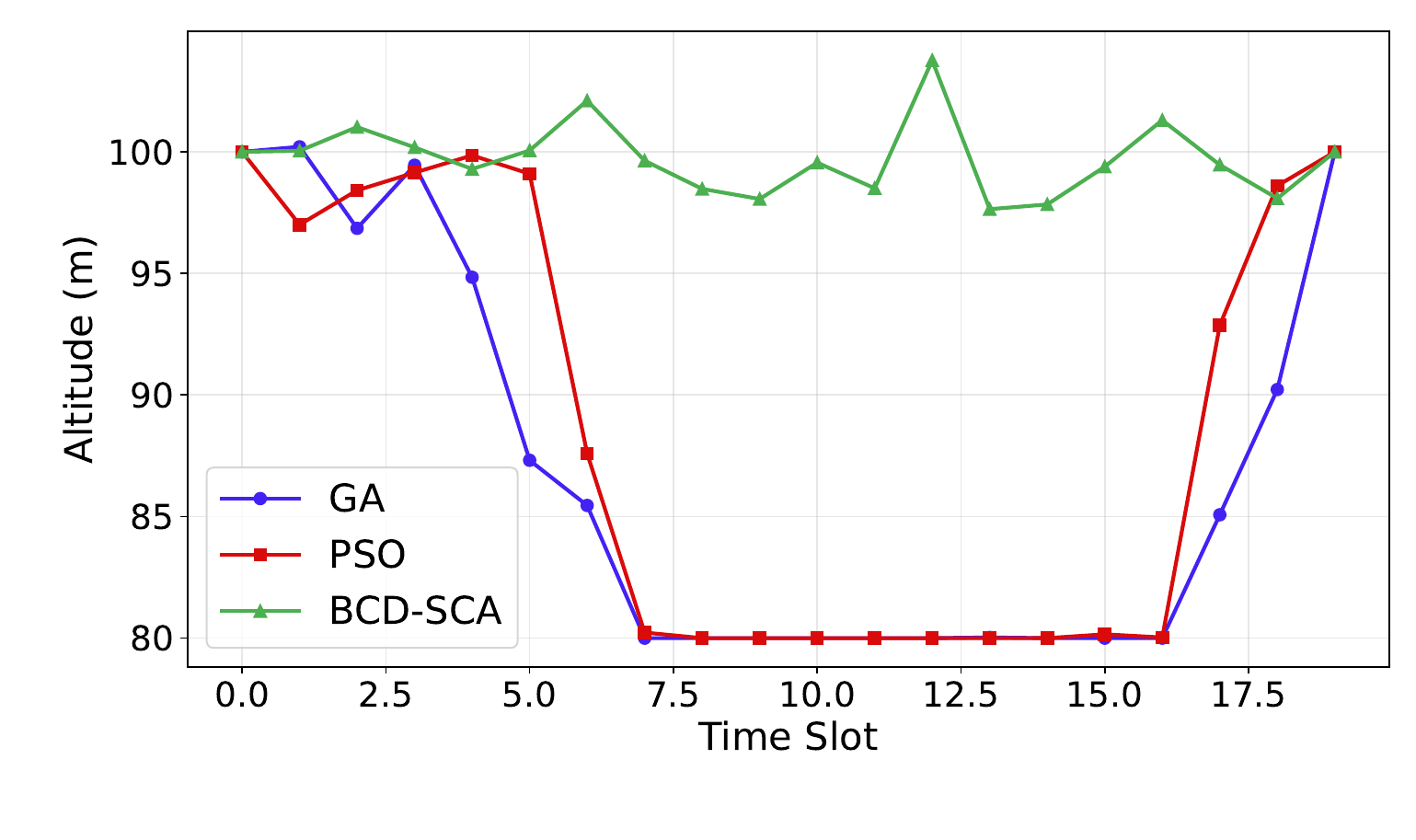}
    \caption{Optimized UAV altitude profiles over time slots.}
    \label{fig:altitude}
\end{figure}
Fig.~\ref{fig:trajectory} and Fig.~\ref{fig:altitude} show the optimized horizontal trajectories and altitude profiles of the UAV. Unlike a straight-line trajectory that ignores the wardens, the considered covertness-aware algorithms adjust the UAV path to avoid high-risk detection regions while maintaining communication with Bob. BCD-SCA obtains a feasible but locally optimal trajectory, whereas GA, PSO provide more favorable paths due to their stronger global search capability. In addition, the altitude profiles indicate that all algorithms exploit 3D trajectory control to balance communication quality, covertness, and energy efficiency.

\begin{figure}[t]
    \centering
    \includegraphics[trim=0cm 1cm 0cm 0.8cm, clip=true,width=0.85\linewidth]{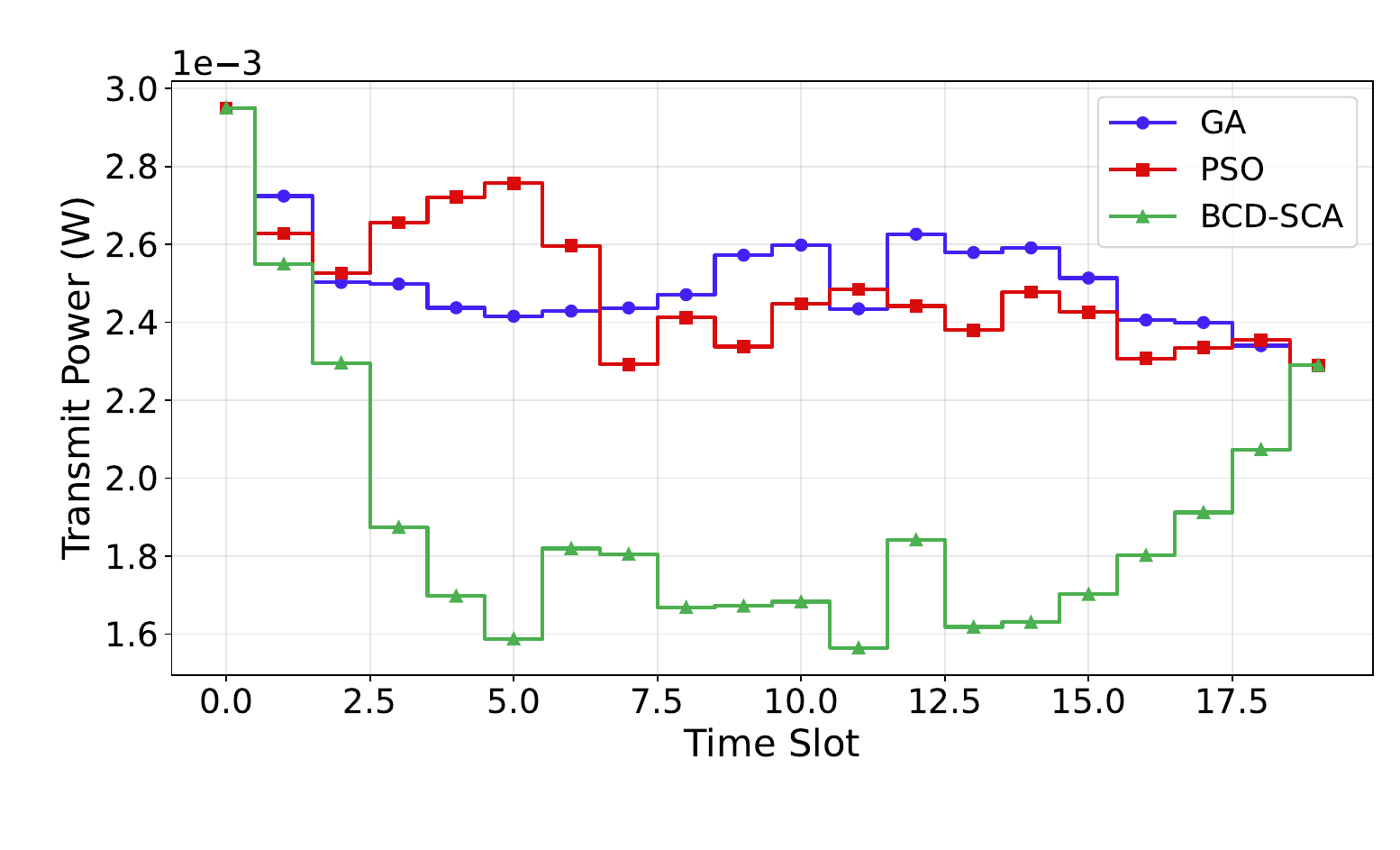}
   \caption{UAV transmit power allocation over time slots.}
    \label{fig:power}
\end{figure}

\begin{figure}[t]
    \centering
    \includegraphics[trim=0cm 1cm 0cm 1.6cm, clip=true, width=0.82\linewidth]{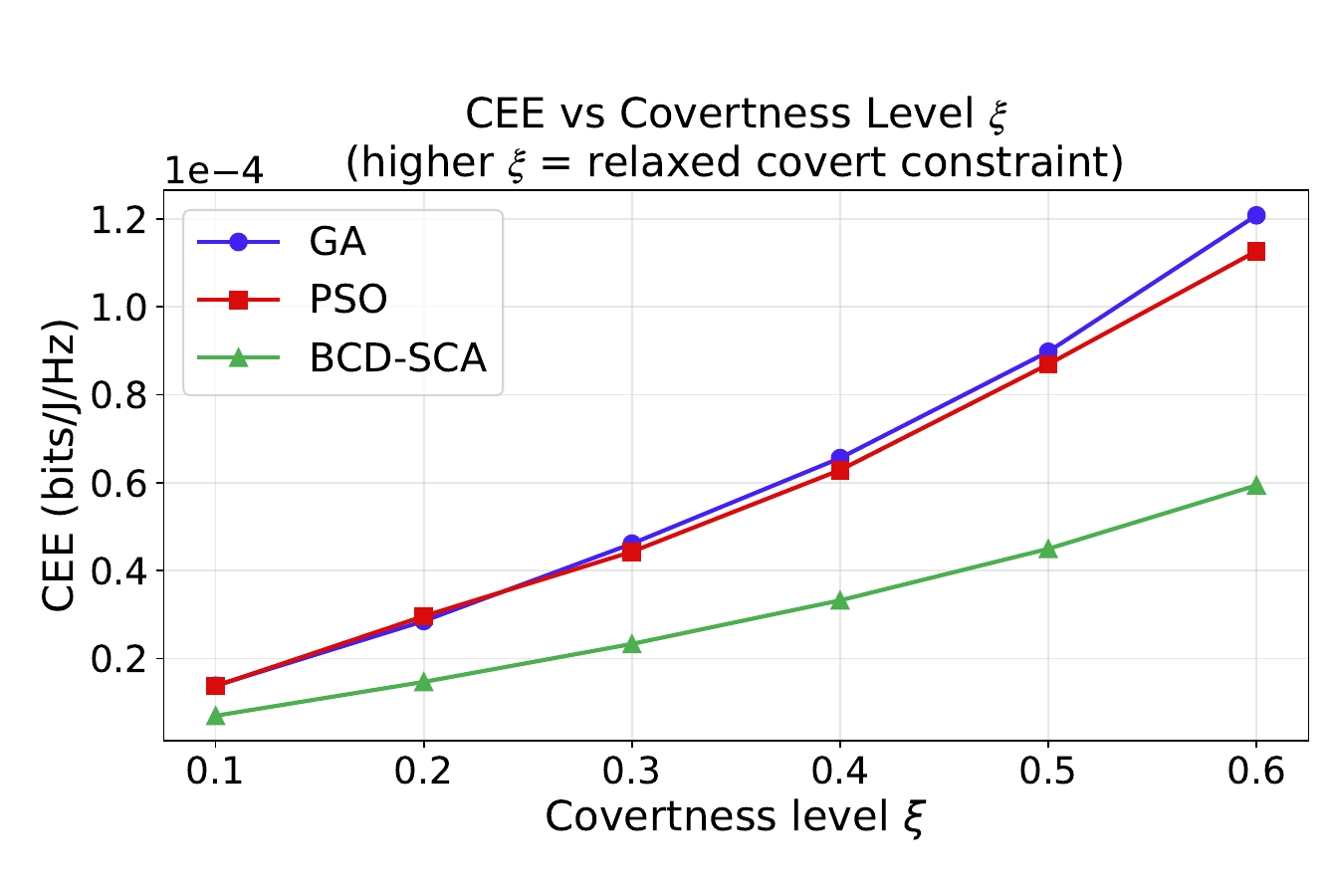}
    \caption{Achievable CEE versus the covertness level.}
    \label{fig:sensitivity_eps}
\end{figure}

\begin{figure}[!ht]
    \centering
    \includegraphics[trim=0cm 0.5cm 0cm 0cm, clip=true, width=0.8\linewidth]{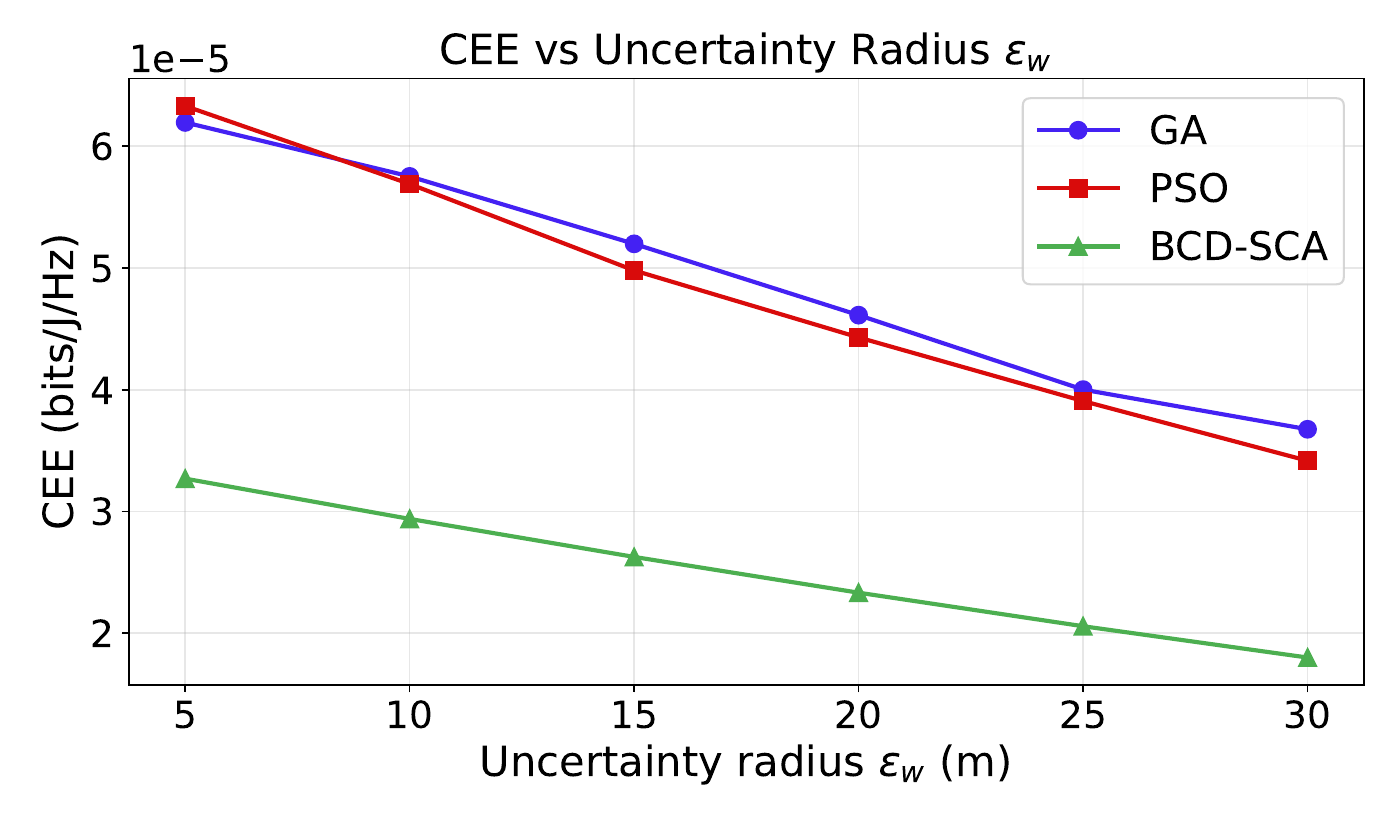}
    \caption{Achievable CEE versus the warden location uncertainty.}
    \label{fig:sensitivity_rk}
\end{figure}

\subsection{Transmit Power Allocation}

Fig.~\ref{fig:power} depicts the UAV's transmit power allocation over the mission time slots. The power allocation highly correlates with the UAV's trajectory. When the UAV is in close proximity to any warden (e.g., around time slots $4-8$ and $12-16$), the transmit power is strictly suppressed to satisfy the stringent covertness constraints and prevent detection. Conversely, when the UAV approaches Bob and moves away from the wardens, the transmit power strategically peaks to maximize the covert throughput. Both the GA and PSO-assisted frameworks demonstrate a more aggressive and effective power allocation strategy compared to the standalone BCD-SCA, contributing to a significantly higher CEE.

\subsection{Impact of System Parameters}

We further evaluate the robustness and performance trade-offs of the considered algorithms. Fig.~\ref{fig:sensitivity_eps} shows that the achievable CEE increases with the covertness level $\epsilon$, since a larger $\epsilon$ relaxes the covertness constraint and allows higher transmit power or more energy-efficient trajectories. Across all covertness levels, GA and PSO outperform the BCD-SCA benchmark, indicating their stronger search capability in the non-convex solution space. Fig.~\ref{fig:sensitivity_rk} shows that increasing the warden location uncertainty radius $r_k$ reduces the CEE, as the UAV must adopt more conservative trajectories and lower transmit power to satisfy worst-case covertness. This result reveals the fundamental trade-off between robust covertness under location uncertainty and covert energy efficiency.
\section{Conclusion}
This paper presented a propulsion-aware and robust optimization framework for UAV-enabled covert communication systems, analyzing both the UAV transmit power and three-dimensional trajectory under strict mobility boundaries and propulsion energy limits. We discussed key security challenges arising from active, colluding wardens under spatial location uncertainty and highlighted the superior performance of our metaheuristic optimization schemes. Numerical results demonstrated that GA and PSO achieve higher CEE than the BCD-SCA benchmark, particularly under stringent covertness and increased location uncertainty. The results confirmed the fundamental trade-off between robustness and energy efficiency.

\section*{Acknowledgment}
This research is funded by Hanoi University of Science
and Technology (HUST) under project number T2025-TN-007.

\bibliographystyle{IEEEtran}
\bibliography{refs}
\end{document}